\documentclass[letterpaper]{article} % DO NOT CHANGE THIS
\usepackage{aaai2026}  % DO NOT CHANGE THIS
\usepackage{times}  % DO NOT CHANGE THIS
\usepackage{helvet}  % DO NOT CHANGE THIS
\usepackage{courier}  % DO NOT CHANGE THIS
\usepackage[hyphens]{url}  % DO NOT CHANGE THIS
\usepackage{graphicx} % DO NOT CHANGE THIS
\usepackage{natbib}  % DO NOT CHANGE THIS AND DO NOT ADD ANY OPTIONS TO IT
\usepackage{caption} % DO NOT CHANGE THIS AND DO NOT ADD ANY OPTIONS TO IT
\usepackage{algorithm}
\usepackage{algorithmic}
\usepackage{amsmath}   
\usepackage{amsfonts} 
\usepackage{amssymb}  
\usepackage{multirow}
\usepackage{booktabs}
\usepackage{xcolor}
\usepackage{pifont}

\usepackage{newfloat}
\usepackage{listings}
\DeclareCaptionStyle{ruled}{labelfont=normalfont,labelsep=colon,strut=off} % DO NOT CHANGE THIS
\floatstyle{ruled}
\newfloat{listing}{tb}{lst}{}
\floatname{listing}{Listing}
\title{SLD-L2S: Hierarchical Subspace Latent Diffusion for High-Fidelity Lip to Speech Synthesis}

\author {
    Yifan Liang\textsuperscript{\rm 1,\rm 2},
    Andong Li \textsuperscript{\rm 1,\rm 2},
    Kang Yang \textsuperscript{\rm 3},
    Guochen Yu \textsuperscript{\rm 4},
    Fangkun Liu \textsuperscript{\rm 1,\rm 2},
    Lingling Dai \textsuperscript{\rm 1,\rm 2},
    Xiaodong Li \textsuperscript{\rm 1,\rm 2},
    Chengshi Zheng \textsuperscript{\rm 1,\rm 2}\thanks{Corresponding author is Chengshi Zheng}
}

\affiliations {
    \textsuperscript{\rm 1}Institute of Acoustics, Chinese Academy of Sciences, Beijing, China
\\
    \textsuperscript{\rm 2}University of Chinese Academy of Sciences, Beijing, China\\
    \textsuperscript{\rm 3}School of Artificial Intelligence and Computer Science, Jiangnan University, Wuxi, China\\
    \textsuperscript{\rm 4}Zhipu AI\\
    liangyifan@mail.ioa.ac.cn, liandong@mail.ioa.ac.cn, 6233110046@stu.jiangnan.edu.cn, guochen.yu@aminer.cn, liufangkun@mail.ioa.ac.cn, dailingling@mail.ioa.ac.cn, lxd@mail.ioa.ac.cn, cszheng@mail.ioa.ac.cn
}

\usepackage{bibentry}
\begin{document}

\maketitle

\begin{abstract}
Although lip-to-speech synthesis (L2S) has achieved significant progress in recent years, current state-of-the-art methods typically rely on intermediate representations such as mel-spectrograms or discrete self-supervised learning (SSL) tokens. The potential of latent diffusion models (LDMs) in this task remains largely unexplored. In this paper, we introduce SLD-L2S, a novel L2S framework built upon a hierarchical subspace latent diffusion model. Our method aims to directly map visual lip movements to the continuous latent space of a pre-trained neural audio codec, thereby avoiding the information loss inherent in traditional intermediate representations. The core of our method is a hierarchical architecture that processes visual representations through multiple parallel subspaces, initiated by a subspace decomposition module. To efficiently enhance interactions within and between these subspaces, we design the diffusion convolution block (DiCB) as our network backbone. Furthermore, we employ a reparameterized flow matching technique to directly generate the target latent vectors. This enables a principled inclusion of speech language model (SLM) and semantic losses during training, moving beyond conventional flow matching objectives and improving synthesized speech quality. Our experiments show that SLD-L2S achieves state-of-the-art generation quality on multiple benchmark datasets, surpassing existing methods in both objective and subjective evaluations.
\end{abstract}

\section{Introduction}

Lip-to-speech (L2S) synthesis is the task of generating audible speech solely from visual lip movements \cite{milner2015}, which presents potential for diverse applications, including automated video dubbing, robust communication in challenging acoustic environments, and assistive technologies for individuals with vocal impairments such as dysphonia and aphonia. The primary challenge in L2S synthesis lies in the ill-posed nature of mapping ambiguous visual cues to the complex acoustic features of speech. Traditional L2S methods \cite{prajwal2020learning,wang2022,kim2024let} are largely confined to constrained settings, characterized by limited vocabularies and speaker-dependent models, limiting their practicality in unconstrained and real-world scenarios.

Recent advances in self-supervised learning (SSL) \cite{hsu2021hubert} and speech synthesis \cite{ren2020fastspeech, polyak2021speech} have catalyzed significant progress in L2S, enabling the generation of high-fidelity speech from corresponding visual inputs. The field has evolved from early, constrained settings \cite{cooke2006audio, harte2015tcd} to robust multi-speaker systems that perform effectively in unconstrained, real-world environments \cite{Chung17, afouras2018lrs3}. The integration of SSL has been particularly transformative, allowing models to learn rich semantic representations from vast unlabeled data. For instance, recent studies \cite{hsu2023,choi2023-intelligible} have leveraged powerful audio-visual SSL models like AV-HuBERT \cite{shi2022learning} to achieve significant gains in the intelligibility of synthesized speech. Concurrently, diffusion-based generative models \cite{defossez2022high,lipman2022flow} have emerged as a powerful paradigm for L2S synthesis, setting a new benchmark for synthesis quality \cite{choi2023,yemini2023lipvoicer,choi2025v2sflow,kim2025faces}.

Despite significant progress, existing L2S methods predominantly rely on generating intermediate representations, such as mel-spectrograms or discrete SSL tokens. These representations, while effective for conveying semantic content, often fail to encapsulate the fine-grained acoustic details requisite for high-fidelity waveform generation. This limitation stems from two primary factors. First, the visual-to-audio mapping is an inherently ill-posed, one-to-many problem: a single lip movement sequence can correspond to multiple valid speech renderings with variations in prosody, emotion, and speaking style. Mel-spectrograms, as a physically-defined acoustic representation, lack the flexibility to model this diversity. Second, while discrete SSL tokens are potent for semantic representation, they are inherently coarse and may discard crucial acoustic details that are necessary for perceptual quality. Other approaches \cite{liang2025naturall2s,liang2025lightl2s} employ end-to-end training, bypassing explicit mapping strategies to prevent potential information loss. However, the resulting speech quality remains limited, suggesting that such methods necessitate high-quality large-scale audio-visual datasets to retrain vocoder. These shortcomings motivates a central question: \textit{Can we identify an alternative representation that more effectively captures the detailed acoustic characteristics required for high-fidelity L2S synthesis?}

Fortunately, latest progress in text-to-speech (TTS) synthesis \cite{wang2023neural,du2024cosyvoice} have demonstrated the effectiveness of neural audio codecs \cite{zeghidour2021soundstream,defossez2022high}, which learn to encode and decode audio waveforms into discrete latent representations. These codecs have shown remarkable performance in generating high-fidelity speech by capturing detailed acoustic characteristics. The success of these codecs in TTS tasks suggests that they are probably be beneficial for L2S synthesis. However, existing L2S methods have not yet fully explored the potential of audio codec latent vectors for L2S. While Uni-Dubing \cite{lei2024uni} has explored incorporating Hifi-Codec \cite{yang2023hifi} to synthesis speech, it frames the problem as predicting multiple residual discrete tokens within a language modeling paradigm. This method presents significant challenges in L2S, because the limited visual input provides insufficient information to support a complicated language modeling task.

In this work, we present SLD-L2S, a novel L2S framework to generate high-fidelity speech directly from visual inputs using a hierarchical subspace latent diffusion model. A key component of our framework is the decomposition of visual representations into multiple subspaces to enable more robust cross-modal mapping. Initially, visual representations are extracted by a pre-trained visual encoder and subsequently processed by our proposed subspace decomposition module. To effectively model these decomposed representations, we introduce diffusion convolution block (DiCB) as the backbone of our generative model. Unlike diffusion transformer (DiT) architectures \cite{peebles2023scalable}, our DiCB leverages the inductive biases of convolution to capture both temporal and cross-subspace dependencies. Furthermore, we employ a reparameterized flow matching \cite{lipman2022flow} objective to enable direct prediction of continuous latents. To enhance speech intelligibility, our training strategy extends beyond a standard flow matching loss  by incorporating two auxiliary loss: a semantic consistency loss imposed on the latent vectors and a speech language model (SLM) loss evaluated on the final synthesized waveform. During the inference, the generated latent vectors are decoded into the final audio waveform by the pre-trained neural audio codec. Leveraging these architectural and methodological innovations, SLD-L2S achieves state-of-the-art results on benchmark datasets. The primary contributions of this work are summarized as follows:

\ding{113} We propose a novel L2S framework for direct generation of the pre-trained audio codec latent vectors from visual inputs via a hierarchical subspace latent diffusion model.

\ding{113} We design diffusion convolution blocks, subspace decomposition and recomposition modules to effectively map different modalities representations in subspaces.

\ding{113} We introduce a reparameterized flow matching technique for direct latent vector generation to improve the generation quality in L2S.

\section{Related Work}
\subsection{Multi-Speaker Lip to Speech Synthesis}
Recent advances in deep learning have transformed L2S synthesis, driving the field from speaker-dependent models in constrained settings toward versatile multi-speaker systems. A pivotal development has been the integration of SSL techniques, which enhances performance by leveraging powerful semantic representations learned from vast unlabeled audio-visual data. For instance, recent studies \cite{hsu2023,choi2023-intelligible,lei2024uni,liang2025naturall2s} have effectively used pre-trained SSL models like AV-Hubert as potent feature extractors, yielding significant improvements in speech intelligibility. Concurrently, generative diffusion models have become a dominant paradigm. Early approaches in this domain, such as DiffV2S \cite{choi2023} and LipVoicer \cite{yemini2023lipvoicer}, employed denoising diffusion probabilistic models (DDPMs) \cite{ho2020denoising} to synthesize mel-spectrograms conditioned on visual features and auxiliary inputs like vision-guided speaker embeddings and semantic information from lip-reading models. More recent works, including V2SFlow \cite{choi2025v2sflow} and From Faces to Voices \cite{kim2025faces} introduce flow matching \cite{lipman2022flow} to further advancing synthesis quality and demonstrating the remarkable potential of diffusion model for L2S.

Except generating intermediate representations like mel-spectrograms or discrete SSL tokens, Uni-Dubbing \cite{lei2024uni} has incorporated Hifi-Codec, which formulates the problem as predicting discrete residual tokens. This aligns with recent trends in TTS, where audio codecs are increasingly used to improve speech quality. However, the information-sparse visual modality provides constraints for the high-precision categorical prediction of discrete tokens. In contrast, our proposed SLD-L2S framework directly generates the latent vectors of a pre-trained audio codec from visual inputs, utilizing a hierarchical subspace latent diffusion model.

\subsection{Neural Audio Codec}
Neural audio codecs are a class of generative models engineered for high-fidelity audio compression \cite{zeghidour2021soundstream,defossez2022high,kumar2023high}. Architecturally, they comprise an encoder that maps continuous raw waveforms into a sequence of discrete tokens, and a corresponding decoder that synthesizes a high-quality waveform from these tokens. The quantization of the output of encoder is typically achieved through residual vector quantization (RVQ). This method recursively quantizes the residual signal from  previous quantization step, with each stage employing a distinct codebook to progressively capture more detail.

Beyond compression, the audio codecs have become instrumental in generative speech tasks. By repurposing the encoder of a pre-trained codec as acoustic tokenizers, speech synthesis can be reframed as a conditional sequence-to-sequence modeling problem, akin to tasks in natural language processing (NLP). A prominent example is VALL-E \cite{wang2023neural}, which auto-regressively predicts discrete acoustic tokens conditioned on textual and acoustic prompts. However, this token-prediction approach is ill-suited for L2S synthesis as previous discussions. Inspired by the approach in NaturalSpeech2 \cite{shen2023naturalspeech}, which models the continuous latent space prior to quantization, we eschew the problematic discrete prediction task. Instead, our method is designed to directly generate the continuous latent vectors of a pre-trained audio codec conditioned on visual inputs.

\begin{figure*}[t]
\centering
\includegraphics[width=1.0\textwidth]{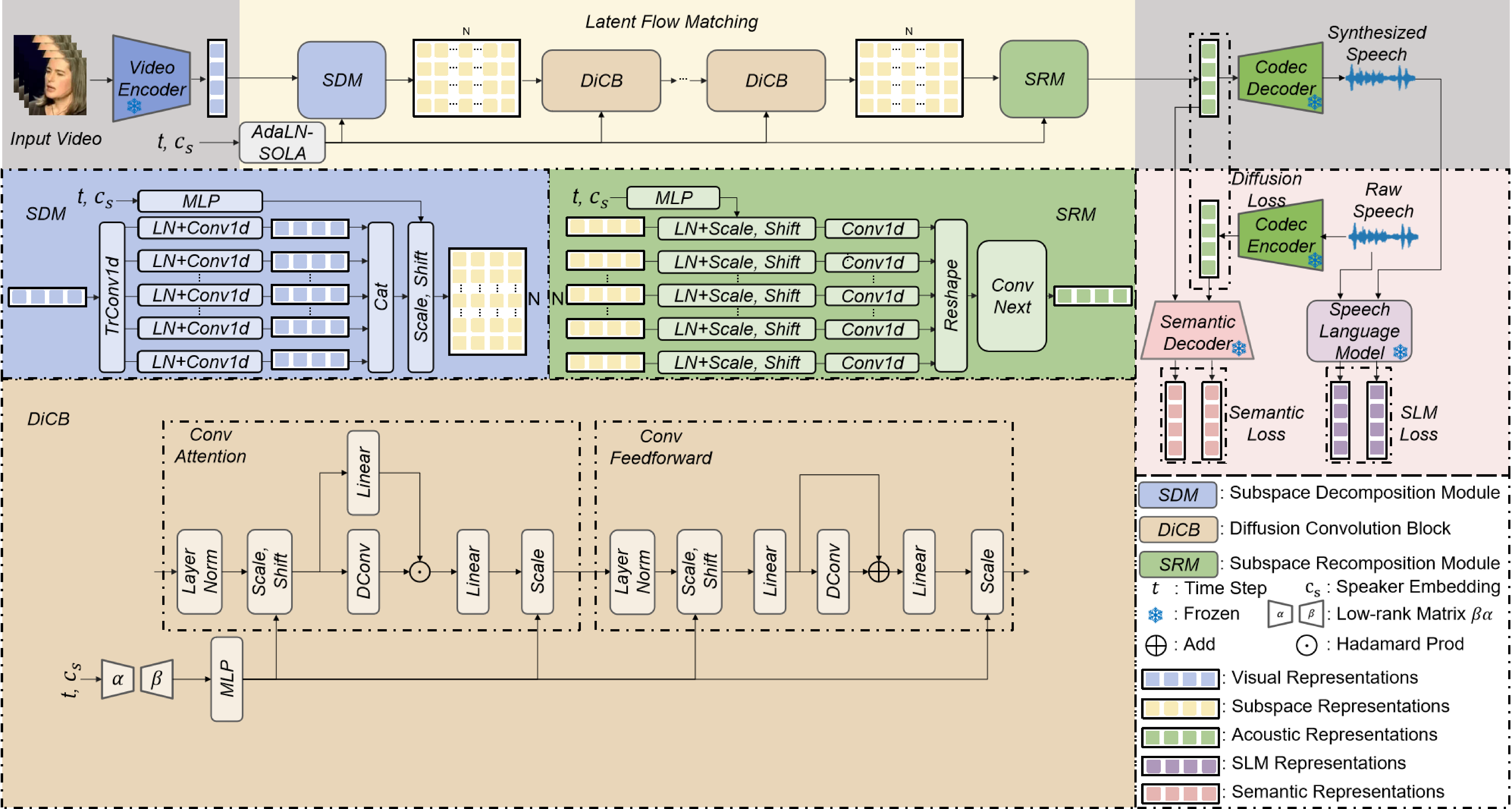}
\caption{The model utilizes a latent flow matching approach to directly map visual features to the continuous acoustic latent space. The architecture consists of three main stages: a subspace decomposition module processes visual features into parallel pathways; a backbone of DiCBs, conditioned with AdaLN-SOLA, effectively refines these representations; and a subspace recomposition module fuses them into a final latent vector. The model is trained with a multi-objective function, including a flow matching loss for acoustic reconstruction, a semantic loss on the latent space, and an SLM loss on the final synthesized waveform. During inference, the model generates the latent vectors for the codec to synthesize into speech.}
\label{fig:overview}
\end{figure*}

\subsection{Flow Matching}
\label{sec:flow_matching}
Flow matching \cite{lipman2022flow} has emerged as a powerful technique for generative modeling, garnering significant attention across diverse domains, including image \cite{esser2024scaling}, video \cite{zheng2024open}, and speech generation \cite{chen2024f5}. Its efficacy in L2S synthesis has been demonstrated by recent works like V2SFLow \cite{choi2025v2sflow} and From Faces to Voices \cite{kim2025faces}. Specifically, V2SFLow achieves speech generation by generating speech through various acoustic attributes, while From Faces to Voices maps these attributes through a hierarchical visual encoder to enable L2S synthesis.

Given samples $x_{1}$ from a target data distribution and samples $x_{0}$ from a standard Gaussian, the core objective of flow matching is to learn a continuous vector field $v_{\theta}(x_t, t)$ that maps $x_0$ to $x_1$ along an ordinary differential equation (ODE) path defined as $dx_{t} = v_{\theta}(x_{t}, t) dt$. A common approach for training involves a simple linear flow, where the path is defined as
\begin{equation}
\label{eq:linear_flow}
x_{t} = (1-t) x_{0} + t x_{1}, t \in [0, 1].
\end{equation}
For this specific linear path, the ground truth velocity field is analytically given by $v(x_{t}, t) = \frac{dx_{t}}{dt} = x_{1} - x_{0}$.
We can train a neural network $v_{\theta}$ to approximate this ground truth velocity at any continuous point $x_t$ on the path from $x_{0}$ to $x_{1}$ at time $t \in [0, 1]$ by minimizing the following objective:
\begin{equation}
\label{eq:flow_matching_loss}
\mathbb{E}_{x_{0} \sim p_0, x_{1} \sim p_1, t} \left[ || v_{\theta}(x_{t}, C, t) - (x_{1} - x_{0}) ||_2^2 \right],
\end{equation}
where $C$ represents the context information guiding the generation process. In our specific case, $C$ corresponds to the visual representations extracted from the input video frames.

During inference, to generate new samples from the target distribution, we start by sampling $x_{0}$ from the source distribution. Then, the sample $x_{t}$ is iteratively updated using the learned velocity function $v_{\theta}(x_{t}, C, t)$ until it approximates $x_{1}$ at the target distribution. This process essentially solves the ODE using numerical methods, commonly the Euler method:
\begin{equation}
x_{t+1} = x_{t} + v_{\theta}(x_{t}, C, t) \Delta t,
\end{equation}
where $\Delta t$ is the step size. This iterative procedure enables sample generation from the target distribution by following the learned velocity field, which effectively guides samples from the source distribution.

\section{Proposed Method}
\subsection{Overview}
The overall architecture of our proposed L2S framework, SLD-L2S, is depicted in Figure 1. The central principle of our approach is to learn a direct mapping from visual representations to the continuous latent space of a pre-trained neural audio codec. To achieve this, we employ a hierarchical subspace latent diffusion model that conditioned on visual representations through multiple parallel subspaces. Except traditional denoising loss, the entire framework is also optimized with a multi-level loss function, designed to ensure both the acoustic fidelity and the semantic consistency of the synthesized speech.

\subsection{Visual Frontend}
The visual frontend of our framework is responsible for extracting linguistic features from the input video sequence. To this end, we employ the visual frontend of AV-Hubert Large \cite{shi2022learning}, a self-supervised audio-visual model pre-trained on the large-scale LRS3 \cite{afouras2018lrs3} and VoxCeleb2 \cite{chung2018voxceleb2} datasets. Consistent with prior work, we find AV-Hubert to be highly effective, as it provides visual representations that are rich in phonetic content, which is well-suited for L2S synthesis.

\subsection{Hierarchical Subspace Latent Flowmatching}
\subsubsection{Subspace Decomposition}
The initial stage of our latent flow matching model involves decomposing the visual representation into a multi-subspace representation. First, the visual features extracted from the AV-Hubert encoder are temporally upsampled to match the resolution of the latent space of target audio codec. Subsequently, instead of a monolithic mapping, the upsampled features are channeled into a set of parallel convolutional subspaces. Each subspace, consisting of layer normalization and 1D convolution, is designed to learn a distinct subspace representation. These parallel subspace representations are then concatenated to form a unified tensor, which serves as the input to the DiCB backbone. Following standard practice in diffusion models \cite{peebles2023scalable}, the time step $t$ and speaker embedding $c_s$ are projected via an adaptive layer norm (AdaLN) block to generate scale and shift parameters for conditional modulation within the network.

\subsubsection{Diffusion Convolution Block}
The mapping from visual to acoustic features is a complex cross-modal transformation characterized by intricate, multi-scale dependencies. While transformer-based architectures like the DiT excel in many generative tasks, their monolithic self-attention mechanism can be sub-optimal for capturing the localized and hierarchical patterns inherent in this problem. To address this, we introduce the DiCB as the fundamental building block of our generative model. Drawing inspiration from both DiT and Conv2Former \cite{hou2024conv2former}, DiCB is designed to efficiently process the decomposed subspace representations.

In a significant departure from DiT, we replace the standard self-attention and MLP layers with specialized convolutional modules, inheriting design principles from Conv2Former. This makes the DiCB architecture inherently insensitive to variations in sequence length, a desirable property for L2S applications. The core of each block is a convolutional attention module. Here, a depthwise convolution with a kernel size of $t \times k$ (where $t$ and $k$ are the kernel dimensions along time and subspace, respectively) is used to model local feature dependencies, and the output is produced via a Hadamard product with the value projection. This design efficiently models local and cross-subspace interactions. This is followed by a convolutional feedforward module, which facilitates channel-wise information mixing, analogous to the FFN in a standard transformer block.

Another key feature of DiCB is its conditioning mechanism. Similar to DiT, we integrate the time step $t$ and speaker embedding $c_s$ using an adaptive layer normalization module. Specifically, we employ AdaLN-SOLA \cite{hai2024ezaudio}, which reduces parameterization and computational overhead by using a single shared AdaLN module across all blocks and timesteps. Furthermore, it incorporates a low-rank adjustment mechanism to modulate the normalization parameters based on the conditioning variables, which we find stabilizes training and preserving performance.

\subsubsection{Subspace Recomposition}
The final stage of our latent flow matching model, the subspace recomposition module, transforms the processed features from the DiCB backbone into the target latent space format of the audio codec. The output of the backbone, which now contains highly integrated feature representations, is first conditioned on the time step $t$ and speaker embedding $c_s$ using an AdaLN layer. To address dimensional mismatch between DiCB outputs and the latent vectors of the audio codec, a 1D convolution and a reshape operation are applied to project the tensor to the correct output dimensions. Finally, we incorporate a ConvNeXt block \cite{liu2022convnet} to ensure a robust fusion of information across the feature channels. This block enables high-order interactions and allows the model to learn an optimal combination of the multi-subspace features.

\begin{table*}[ht]
    \renewcommand{\arraystretch}{1.18}
    \renewcommand{\tabcolsep}{1.0mm}
    \centering
    {
    \begin{tabular}{l c ccccc ccccc}
        \toprule
        \multirow{2}{*}{\textbf{Method}} & \multirow{2}{*}{\textbf{NFE}} & \multicolumn{5}{c}{\textbf{LRS3-TED}} & \multicolumn{5}{c}{\textbf{LRS2-BBC}} \\
        \cmidrule(l{2pt}r{2pt}){3-7} \cmidrule(l{2pt}r{2pt}){8-12}
        & & UTMOS$\uparrow$ & SCOREQ$\uparrow$ & D-BERT$\uparrow$ & WER$\downarrow$ & SECS$\uparrow$  & UTMOS$\uparrow$ & SCOREQ$\uparrow$ & D-BERT$\uparrow$ & WER$\downarrow$ & SECS$\uparrow$  \\
        \cmidrule(l{2pt}r{2pt}){1-2} \cmidrule(l{2pt}r{2pt}){3-7} \cmidrule(l{2pt}r{2pt}){8-12} 
        Ground Truth & -- & 3.5721 & 3.8105 & -- & 0.71 & -- & 3.0538 & 3.3008 & -- & 1.35 & --  \\ 
        Intelligible & 1 & 2.7019 & 2.7024 & 0.8231 & 27.73 & 0.761$^*$ & 2.3315 & 2.3437 & 0.7932 & 36.50 & 0.701$^*$ \\
        DiffV2S & 1000 & 3.0681 & 3.3811 & 0.7408 & 35.81 & 0.627~~ & 2.9575 & 3.2721 & 0.7103 & 47.58 & 0.569~~ \\
        LipVoicer & 400 & 2.4540 & 2.6598 & 0.7132 & \textbf{20.62} & 0.588~~ & 2.3756 & 2.7973 & 0.6906 & \textbf{24.92} & 0.578~~ \\
        NaturalL2S & 1 & 3.6598 & 3.7695 & 0.8068 & 30.37 & 0.628~~ & 3.5014 & 3.6702 & 0.7838 & 39.19 & 0.562~~ \\
        V2SFlow & 30 & 3.6939 & 4.0710 & \textbf{0.8306} & 27.55 & \textbf{0.851}$^*$ & 3.4556 & 3.9157 & \textbf{0.8064} & 34.17 & \textbf{0.820}$^*$ \\
        \textbf{Proposed} & 10 & \textbf{4.2096} & \textbf{4.6075} & 0.8284 & 30.22 & 0.804$^*$ & \textbf{4.1664} & \textbf{4.5706} & 0.8005 & 39.54 & 0.756$^*$ \\
        \bottomrule
    \end{tabular}
    }
    \caption{
        Lip-to-speech synthesis performance comparisons on LRS3-TED and LRS2-BBC dataset.
        $\uparrow$: higher is better, $\downarrow$: lower is better. The $^*$ indicates methods utilizing reference speaker embedding.
    }
    \label{table:1}
\end{table*}

\subsection{Training Objectives}
\subsubsection{Reparameterized Flow Matching Loss}
The standard Flow Matching objective seeks to estimate the velocity field $v_{\theta}$ by minimizing a mean squared error (MSE) loss, as defined in Eq.~\ref{eq:flow_matching_loss}. However, this formulation can be unstable in our practice because it requires the network to regress a noise-dependent random derivative. The stochastic nature of this target prevents the use of auxiliary losses and complicates optimization. Inspired by \cite{fu2025moflow}, we adopt a reparameterized objective. Instead of directly predicting the velocity field $v_{\theta}$, we reformulate the output of our model to be a direct estimate of the target data point, $x_1$. This reparameterization not only improves training stability but also provides a more intuitive framework for incorporating auxiliary losses that operate in the data space. Specifically, we define the output of our model $d_{\theta}(x_{t}, c_{s}, t)$ as a direct prediction of the target sample $x_1$. This relates to the velocity field $v_{\theta}$ via the linear probability path defined in Eq.~\ref{eq:linear_flow}:
\begin{equation}
d_{\theta}(x_{t}, c_{s}, t) = x_{t} + (1-t)v_{\theta}(x_{t}, c_{s}, t).
\label{eq:reparameterized_output}
\end{equation}

Consequently, the velocity field $v_{\theta}$ is learned implicitly. The reparameterized Flow Matching loss, $\mathcal{L}_{FM}$, is then formulated as a weighted MSE between the model's prediction and the target $x_{1}$:
\begin{equation}
\mathcal{L}_{FM} = \mathbb{E}_{x_{0}, x_{1}, t} \left[ \frac{|| d_{\theta}(x_{t}, c_{s}, t) - x_{1} ||_2^{2}}{(1-t)^2} \right],
\label{eq:reparameterized_flow_matching_loss}
\end{equation}
where $x_0 \sim p_0$, $x_1 \sim p_1$, and $t \sim U(0,1)$. 

\subsubsection{Auxiliary Losses}
To improve the perceptual quality and content accuracy of the synthesized audio, we first incorporate SLM loss based on a pre-trained SLM. This SLM loss facilitates knowledge transfer by enforcing consistency in a high-level perceptual feature space. Specifically, we decode the predicted latent vector $d_{\theta}(x_{t}, c_{s}, t)$ and the target latent vector $x_{1}$ into waveforms using the audio codec's decoder $D_{a}(\cdot)$. We then extract feature representations from both waveforms using the SLM and compute the squared L2 distance between them:
\begin{equation}
\begin{split}
\mathcal{L}_{SLM} = & 
\mathbb{E}_{x_{0}, x_{1}, t} \left[ \left\| SLM\left(D_{a}\left(d_{\theta}(x_{t}, c_{s}, t)\right)\right) - \right. \right. \\ & \left. \left. SLM\left(D_{a}(x_{1})\right) \right\|_2^{2} \right],
\end{split}
\label{eq:slm_loss}
\end{equation}

Meanwhile, inspired by recent work on semantically-aware audio codecs like X-codec \cite{ye2025codec}, we introduce a semantic loss to enforce content consistency directly at the latent level. This loss ensures that the generated codec latents retain the same core semantic content as the target latents. To achieve this, we use a separately pre-trained semantic decoder, $D_{s}(\cdot)$, which is trained in an auto-encoder setup (with a corresponding semantic encoder) to reconstruct semantic features (e.g., from HuBERT \cite{hsu2021hubert} or WavLM \cite{chen2022wavlm}) from the codec latents. The semantic loss is the squared L2 distance between the reconstructed semantic features of the predicted and target latents:
\begin{equation}
\mathcal{L}_{sem} = \mathbb{E}_{x_{0}, x_{1}, t} \left[ || D_{s}(d_{\theta}(x_{t}, c_{s}, t)) - D_{s}(x_{1}) ||_2^{2} \right],
\label{eq:semantic_loss}
\end{equation}
where $D_{s}(\cdot)$ is the pre-trained semantic decoder, which is implemented as a multi-layer convolutional network.

The final training objective for our SLD-L2S framework is a weighted combination of these multi-objective losses:
\begin{equation}
\mathcal{L} = \mathcal{L}_{FM} + \lambda_1 \mathcal{L}_{SLM} + \lambda_2 \mathcal{L}_{sem},
\end{equation}
where $\lambda_1$ and $\lambda_2$ are hyperparameters balancing the contribution of each auxiliary objective. This multi-objective loss structure allows the model to jointly optimize for perceptual quality and semantic content, leading to improvements in overall synthesis performance.

\section{Experimental Setup}
\subsection{Datasets}
Our experiments are conducted on multi-speaker audio-visual benchmarks: LRS3-TED \cite{afouras2018lrs3} and LRS2-BBC \cite{Chung17}. The LRS3-TED dataset, derived from TED and TEDx talks, contains approximately 439 hours of footage from over 5,000 speakers across 150,000 utterances. The LRS2-BBC dataset consists of 200 hours of data from BBC broadcasts, comprising 144,000 segments from a diverse set of speakers. We only use the LRS3-TED corpus for training, following its official pre-defined splits for training, validation, and testing. To assess the generalization capabilities of each method, we also evaluate on the LRS2-BBC test set.

\subsection{Implementation Details}
Our data pre-processing follows the pipeline in \cite{shi2022learning}. For each video, we extract facial landmarks using the dlib \cite{king2009dlib} library, crop the mouth region-of-interest (ROI), resize it to an 88x88 resolution and convert to grayscale. These frames are then processed by a pre-trained AV-Hubert Large model, which functions as our visual encoder, yielding 1024-dimensional feature vectors. For audio synthesis, we employ X-Codec-hubert as our neural audio codec, which operates at a 16 kHz sampling rate and was pre-trained on the LibriSpeech corpus. The speaker identity is provided via a 256-dimensional embedding $c_s$, extracted from a reference utterance using a GE2E \cite{wan2018generalized}. For our auxiliary losses, semantic features are extracted using a pre-trained HuBERT-base-ls960 model and we use WavLM \cite{chen2022wavlm} for calculating SLM loss. To align the temporal resolution mismatch between the video and the audio, we upsample the visual features by using a transposed convolution layer.

The architecture of our SLD-L2S model is configured as follows. The subspace decomposition module partitions the 1024-dimensional visual features into 8 subspaces. Each subspace, containing a layer normalization and a 1D convolution, processes the features before they are concatenated. The backbone is composed of 12 DiCBs. Within each DiCB, the convolutional attention module employs a kernel of size (5, 7) for the time and subspace dimensions, respectively. The convolutional feedforward module uses a kernel size of 3 and expands the channel dimension to 1024. Conditional inputs are integrated via AdaLN-SOLA module, with the low-rank adjustment matrix rank set to 32. Finally, the subspace recomposition module projects the output from each of the 8 subspace streams to a dimension of 128. These are then concatenated and fused into a single 1024-dimensional vector. This fused representation is passed through ConvNeXt blocks, which consists of 3 layers with 512 hidden channels to facilitate final cross-channel interactions before producing the output. For training, we use the AdamW optimizer initialized with a learning rate of 2e-4 and the cosine learning rate decay strategy. The batch size is set to 16, and the model is trained for 150K iterations on a single NVIDIA H100 GPU over about five days. The AdamW scheme is configured with $\beta_1 = 0.8$, $\beta_2 = 0.99$ The hyperparameters for the auxiliary losses are set to $\lambda_1 = 1$ and $\lambda_2 = 100$.

\begin{table}[t]
\centering
\setlength{\tabcolsep}{1mm}
\footnotesize
\begin{tabular}{lccc}
\toprule
Method & Naturalness$\uparrow$ & Intelligibility$\uparrow$ & Similarity$\uparrow$ \\ \cmidrule(lr){1-4}
Ground Truth & 4.36 $\scriptstyle{\pm 0.15}$ & 4.48 $\scriptstyle{\pm 0.15}$ & -- \\
Intelligible & 2.82 $\scriptstyle{\pm 0.25}$ & 3.37 $\scriptstyle{\pm 0.26}$ & 3.02 $\scriptstyle{\pm 0.17}$ \\
DiffV2S      & 2.95 $\scriptstyle{\pm 0.27}$ & 2.98 $\scriptstyle{\pm 0.27}$ & 2.80 $\scriptstyle{\pm 0.23}$ \\
LipVoicer    & 2.51 $\scriptstyle{\pm 0.23}$ & 2.97 $\scriptstyle{\pm 0.29}$ & 2.59 $\scriptstyle{\pm 0.23}$ \\
NaturalL2S   & 3.54 $\scriptstyle{\pm 0.32}$ & 3.43 $\scriptstyle{\pm 0.27}$ & 3.21 $\scriptstyle{\pm 0.21}$ \\
V2SFLow      & 3.88 $\scriptstyle{\pm 0.25}$ & 3.69 $\scriptstyle{\pm 0.25}$ & 3.90 $\scriptstyle{\pm 0.14}$ \\
Proposed     & 4.17 $\scriptstyle{\pm 0.24}$ & 3.65 $\scriptstyle{\pm 0.25}$ & 3.77 $\scriptstyle{\pm 0.15}$ \\
\bottomrule
\end{tabular}
\caption{MOS results on LRS3 dataset.}
\label{table:3}
\end{table}

\subsection{Evaluation Metrics}
We perform a comprehensive evaluation of our proposed SLD-L2S framework using a suite of objective and subjective metrics. The evaluation is designed to assess three critical aspects of the synthesized speech: perceptual quality, content intelligibility, and speaker similarity. For objective evaluation, we employ the following metrics:

\noindent \ding{113} \textbf{Quality}: We assess perceptual quality using UTMOS \cite{saeki2022utmos} and SCOREQ \cite{ragano2024scoreq}. Both are non-intrusive, reference-free models that predict a mean opinion score (MOS) on a 1-to-5 scale, designed to correlate with human quality judgments.

\noindent \ding{113} \textbf{Intelligibility}: We measure intelligibility from two perspectives. First, the word error rate (WER) is calculated using the AUTO-AVSR \cite{ma2023auto}. Second, we utilize SpeechBERTScore (D-BERT) \cite{saeki2024speechbertscore}, which leverages NLP evaluation metrics. It calculates the cosine similarity between the SSL embeddings of the synthesized and target speech, providing a semantic similarity score in the range of –1 to 1.

\noindent \ding{113} \textbf{Speaker Similarity}: We quantify speaker identity preservation using speaker embedding cosine similarity (SECS). This metric is calculated from speaker embeddings, which are extracted from both the synthesized and target speech using the Resemblyzer \cite{wan2018generalized} library.

Subjective evaluation is conducted through a MOS test. A group of 15 participants listened to 30 synthesized samples and were asked to rate each clip on three criteria: naturalness (how natural the speech sounds), intelligibility (how easy the content is to understand), and speaker similarity (how similar the synthesized speech is to the target speaker). All criteria were rated with 5 points (from 1 to 5).

\begin{table}[t]
\centering
\setlength{\tabcolsep}{1mm}
\footnotesize
\begin{tabular}{lccccc}
\toprule
Method 
      & UTMOS$\uparrow$ & SCOREQ$\uparrow$ & D-BERT$\uparrow$ & WER$\downarrow$ & SECS$\uparrow$     \\ \cmidrule(lr){1-6}
\textbf{Ours} & 4.2096 & 4.6075 & 0.8284 & 30.22 & 0.804  \\ \cmidrule(lr){1-6}
\emph{w/o} DiCB  & 4.0835 & 4.5056 & 0.8184 & 30.70 & 0.618  \\  
\emph{w/o} Repara   & 3.9540 & 4.2600 & 0.8249 & 29.70 & 0.809  \\ 
\emph{w/o} $L_{sem}$   & 4.1859 & 4.5830 & 0.8272  & 30.21 & 0.795  \\
\emph{w/o} $L_{SLM}$  & 4.2885 & 4.6958 & 0.8207  & 29.99 & 0.776  \\
\emph{~~w/o} $L_{sem}$   & 4.2507 & 4.6494 & 0.8157  & 30.53 & 0.771\\
\bottomrule
\end{tabular}
\caption{Ablation study results on the LRS3 dataset. Repara indicates the inclusion of our reparameterized flow matching object. DiCB signifies the replacement of our proposed DiCB backbone with DiT, maintaining closely matched parameter sizes (DiCB: 57.7M, DiT: 57.8M).}
\label{table:ablation}
\end{table}

\section{Results and Discussion}
\subsection{Comparison with State-of-the-Art Methods}
We compare our proposed SLD-L2S framework against advanced L2S methods: Intelligible \cite{choi2023-intelligible}, DiffV2S \cite{choi2023}, LipVoicer \cite{yemini2023lipvoicer}, NaturalL2S \cite{liang2025naturall2s}, and V2SFlow \cite{choi2025v2sflow}. The results on the LRS3 and LRS2 datasets are presented in Table \ref{table:1}. Our analysis focuses on key aspects including perceptual quality, efficiency, intelligibility and speaker similarity. Notably, the LRS2 dataset is not used during training in order to evaluate the robustness.

First, the results unequivocally establish SLD-L2S as the new state-of-the-art in perceptual quality. As shown in Table \ref{table:1}, our model substantially outperforms all baselines on both datasets in the reference-free quality metrics. On the primary LRS3-TED benchmark, for instance, SLD-L2S surpasses the strongest flow matching based competitor, V2SFlow, by a margin of 0.5157 in UTMOS. Crucially, this superior synthesis quality is achieved with exceptional computational efficiency. SLD-L2S requires only 10 number of function evaluations (NFE) at inference, representing higher efficiency than other diffusion-based methods.

In terms of intelligibility, LipVoicer achieves the lowest WER because its reliance on a powerful, pre-trained visual speech recognition (VSR) model \cite{ma2023auto} to extract semantic features. However, this comes at the cost of degraded speech quality, as indicated by its low UTMOS and SCOREQ scores. Moreover, the require of a VSR model limits its applicability to scenarios where text transcripts are unavailable. In contrast, SLD-L2S strikes a superior balance. It generates intelligible speech, evidenced by its competitive D-BERT scores, which outperforms nearly all the methods without using specialized VSR front-ends and is highly competitive with the top performer, V2SFlow. This indicates that our model effectively generates intelligible speech while simultaneously preserving its perceptual quality.

For speaker identity preservation, V2SFlow achieves the highest SECS score as its training objective explicitly includes a speaker embedding loss to enforce speaker similarity. Despite not employing such a direct loss, our framework achieves a highly competitive second-best SECS score. This result underscores the effectiveness of our proposed DiCB module in conditioning the generation process on the speaker identity without requiring direct supervision.

The results of subjective evaluation, presented in Table 2, further validate the effectiveness of our framework. Our model excels in naturalness, achieving an outstanding score of 4.17 that notably surpasses all competitors and closely approaches the ground truth. In terms of intelligibility, our method ranks second with a score of 3.65. Focusing on speaker similarity, our model obtains a highly competitive score of 3.77, nearly matching the top-performing V2SFlow. This evaluation also highlights the limitations of relying solely on objective metrics. For instance, LipVoicer performs well on the objective WER metric but scores poorly in subjective tests, indicating that its poor acoustic quality and unnaturalness can severely degrade the practical intelligibility for human listeners. This discrepancy underscores that perceptual quality is a critical goal for L2S systems, as it directly impacts the ability of humans to comfortably understand the generated speech.

\begin{table}[t]
\centering
\setlength{\tabcolsep}{1mm}
\footnotesize
\begin{tabular}{lccccc}
\toprule
Kernel size 
      & UTMOS$\uparrow$ & SCOREQ$\uparrow$ & D-BERT$\uparrow$ & WER$\downarrow$ & SECS$\uparrow$     \\ \cmidrule(lr){1-6}
\textbf{Ours (5$\times$7)}  & 4.2096 & 4.6075 & 0.8284 & 30.22 & 0.804 \\ \cmidrule(lr){1-6}
3$\times$5  & 4.2120 & 4.6016 & 0.8292 & 30.34 & 0.801 \\
7$\times$9  & 4.1914 & 4.5733 & 0.8291 & 30.17 & 0.800 \\ 
9$\times$11 & 4.2008 & 4.5852 & 0.8288 & 30.10 & 0.803 \\  
\bottomrule
\end{tabular}
\caption{Ablation study results of different kernel size on the LRS3 dataset. The kernel size is denoted as (time $\times$ subspace).}
\label{table:kernel}
\end{table}

\subsection{Ablation Study}
To analyze the contribution of each component, we conduct an ablation study with results presented in Table \ref{table:ablation}.

\noindent \ding{113} \textbf{DiCB Architecture:} Replacing our DiCB backbone with a parameter-matched DiT leads to a significant performance collapse across nearly all metrics. This result underscores the critical role of our DiCB design in achieving high-quality synthesis and effective speaker conditioning.

\noindent \ding{113} \textbf{Reparameterized Training Objective:} Reverting to the standard velocity-prediction flow matching objective results in a notable drop in perceptual quality (UTMOS, SCOREQ). This confirms that our reparameterized approach is crucial for improving training stability and synthesis quality.

\noindent\ding{113} \textbf{Auxiliary Losses:}
Removing the semantic loss (w/o $\mathcal{L}_{sem}$) causes a minor drop in quality and speaker similarity, suggesting its role as a helpful regularizer for the latent space. More interestingly, removing the SLM loss (w/o $\mathcal{L}_{SLM}$) reveals a trade-off: while this slightly improves the acoustic quality metrics (UTMOS/SCOREQ), it degrades both intelligibility (D-BERT) and speaker similarity (SECS). This indicates that the SLM loss is crucial for content and identity preservation, even if it introduces a slight constraint on the acoustic non-intrusive metrics. Removing both losses leads to a more pronounced degradation, confirming their combined importance.

\begin{table}[t]
\centering
\setlength{\tabcolsep}{1mm}
\footnotesize
\begin{tabular}{lccccc}
\toprule
Subspaces 
      & UTMOS$\uparrow$ & SCOREQ$\uparrow$ & D-BERT$\uparrow$ & WER$\downarrow$ & SECS$\uparrow$     \\ \cmidrule(lr){1-6}
\textbf{Ours (8)}  & 4.2096 & 4.6075 & 0.8284 & 30.22 & 0.804 \\ \cmidrule(lr){1-6}
4  & 4.1797 & 4.5638 & 0.8278 & 29.94 & 0.802 \\
16  & 4.1752 & 4.5683 & 0.8295 & 29.92 & 0.802 \\ 
32 & 4.1288 & 4.5219 & 0.8268 & 30.35 & 0.799 \\  
\bottomrule
\end{tabular}
\caption{Ablation study results of different number of subspaces on the LRS3 dataset.}
\label{table:subspace}
\end{table}

\subsection{Analysis on the design choices of DiCB}
We further ablate two key design choices for our DiCB module: the convolutional kernel size and the number of subspaces. The results are summarized below.

\noindent \ding{113} \textbf{Kernel Size.} As shown in Table \ref{table:kernel}, model performance is remarkably robust to the kernel size in the convolutional attention module in DiCB. In contrast to the trend of using large kernels in computer vision, smaller kernels perform competitively in our task. The 5$\times$7 kernel yields the best balance across all metrics, validating our choice.

\noindent \ding{113} \textbf{Number of Subspaces.} This proves to be a more sensitive hyperparameter. Table \ref{table:subspace} reveals a distinct performance peak at our chosen configuration of 8 subspaces. Using fewer (4) or more (16, 32) subspaces leads to a marked degradation in perceptual quality (UTMOS/SCOREQ). This suggests that 8 subspaces achieve an optimal trade-off, effectively decomposing visual features without creating overly fragmented representations that are difficult to integrate.
\section{Conclusion}
In this paper, we introduced SLD-L2S, a novel framework that sets a state-of-the-art in lip-to-speech synthesis by directly mapping visual features to the latent space of audio codec. Our method leverages an efficient, reparameterized flow matching technique to train a hierarchical subspace network. This network is built upon our proposed diffusion convolution block, which effectively captures the intricate relationships between visual and auditory modalities. Combined with a dual semantic and SLM loss strategy for enhanced content fidelity, SLD-L2S notably surpasses existing methods in perceptual quality. Our work validates the potential of latent diffusion in lip-to-speech synthesis and provides a robust blueprint for the future of high-fidelity L2S synthesis.

\bibliography{aaai2026}

\end{document}